# Photoluminescence, amplification and generation of optical media in without inversion and of induced radiation.

Federal State Institution of Science Prokhorovs Institute of General Physics
Russian Academy of Sciences (GPI RAS).
Russian Federation, 119991, Moscow, 38, Vavilovs st.
ogluzdin@kapella.gpi.ru

The observation of photoluminescence (PL) or achieve lasing modes in alcohol solutions of dyes or other media (silica glass fibers), doped in small quantities by certain active working elements, which are dipped in the environment, do not requires start the inversion, since the main role in time of the optical pumping process is given four-photon interaction and the associated amplification of those frequency components, that result from this interaction, and participating in the PL or generation.

V.E.Ogluzdin

Federal State Institution of Science Prokhorovs Institute of General Physics
Russian Academy of Sciences (GPI RAS)
Russian Federation, 119991, Moscow, 38, Vavilovs st.
ogluzdin@kapella.gpi.ru

It is considered that a necessary condition for achieving laser action is to create the inverted population, which implementation-launch (early) electrons to high-lying energy levels of atoms (molecules) of the active media. This is indeed true in the case of, for example, an electrical discharge in gas mixtures. This fact allows the use of stimulated emission for the action of helium-neon, argon, nitrogen and other lasers, the excitation (inversion) of the working environment that are or electrical discharge or electron beam.

However, to obtain lasing or just PL alcoholic solutions of dyes or other media (quartz glass, optical fibers), doped in small quantities by certain elements (erbium, holmium, neodymium, bismuth), the need to fulfill the above condition is not evident. The observation of photoluminescence or achieve lasing modes in these environments is not envisaged in the working environment, start the inversion, since the main role belongs to the resonant (almost resonant) four-photon interaction processes and the associated process to strengthen one of the emerging among the frequency components that result from these interactions. Note that the four-photon interaction in a natural way is directly related to the law of conservation of energy [1-4]:

$$h\nu + h\nu = h\nu_{pl} + h\nu_{ik}, \quad (1)$$

$\nu$ - frequency of the pump, $\nu_{pl}$ - one of the frequencies components of the emission PL (or oscillation frequency), $\nu_{ik}$ - frequencies components of the dipole transitions accompanied photoluminescence (generation) - the Bohr frequency.

After canceling h, Planck's constant, the ratio (1) corresponds to the definition presented by the phenomenon of photoluminescence: the frequency of the pump is the average arithmetic between the frequency of the peak of the photoluminescence spectrum and the Bohr frequency of the electronic transition responsible for the frequency of this peak photoluminescence [1-2]:

$$2\nu = \nu_{pl} + \nu_{ik}. \quad (2)$$

This definition corresponds to the well-known rule on mirror symmetry of the absorption and PL-spectrum.

The analysis shows that the PL or generation (when a medium placed in the cavity of resonator), in a medium, excited by light emission pumping, are accompanied by the transitions of electrons of atoms of elements, which are doped the host material, to higher energy levels, compared with the start of their position. In accordance with the law of conservation energy (1) these transitions closes the process of photoluminescence.

In fact, in the doping atom, suspended in an isotropic homogeneous medium (host material), in the elementary act of photoluminescence electrons transfer from the level of «i» to the level «k» due to the addition of two quanta of the pump radiation, and the emission at the difference frequency of one photon photoluminescence (or generated laser radiation). For PL quantum efficacy is limited by 50%.

In an environment (host material) in which the process of PL excited or generation of optical radiation, usually there is a rise in temperature. The increase in temperature is associated with relaxation processes. Usually relaxation is nonradiative process and lead to heating of the medium.

The problem of the heat has always been the actual in the development of dye lasers, as well as in solids.

Let us to see some examples to illustrate the preparation of the photoluminescence and generation in the environments in which the photo luminescent (generation), the processes does not require the creation of inversion, and due to interaction of the four lights quantum in accordance with (1).

A) The peak of the photoluminescence spectrum and lasing atomic holmium, placed in a quartz glass fiber corresponds to a wavelength of 2.1 microns [5,6].
The excitation source had wavelength 1.147 nm. There are electronic transitions of the «i» → «k», responsible for the photoluminescence and lasing in the 2.1 micron spectral range; for such an excitation source, according to the relation (1) corresponds to the table value of the wavelength 755.09 nm [4]. In atomic holmium it is an electronic transition 5 I (9 - $^{55}/_2$) - 4 I ($^{15}/_2$ - $^{9}/_2$).

B) About strengthening the light in the glass, activated with erbium [7]. According to the above concept for maximum gain at a wavelength of 1530 nm in the case of a pump wavelength of 980 nm is responsible electronic transition of atomic erbium, which corresponds to the tabulated value of the wavelength 719.7 nm.

C) In the past there were many reports of photoluminescence glass activated bismuth, and the generation of activated bismuth obtained fibers. It remains on the results we agree on PL, published in [8], and according to (1) identify lines of atomic bismuth responsible for the Bohr-frequencies ($v_{ik}$) [4].

| Table value wavelength atomic lines bismuth [9], corresponding to Bohr frequencies ($v_{ik}$) for: | Wavelength of pumping: | Wavelength max. peak the PL spectra: |
|---|---|---|
| 1) 306.722 nm | 502 nm | 1153.5 nm |
| 2) 472.237 nm | 680 nm | 1085.4 nm |
| 3) 552.2 nm | 738 nm | 1171.6 nm |

Note that the spacing between the line, responsible for photoluminescence, and the wavelength of the peak in the photoluminescence spectrum can reach very high values (about 800 nm).
It is also interesting to consider the use of this model for the most studied case of neodymium activation glass. In this case it is necessary consider the electronic transitions in atomic neodymium, not involving ionic structure. It is interesting for laser pump.

Conclusion.

The observation of PL or achieve lasing modes in alcohol solutions of dyes or other media (silica glass fibers), doped in small quantities by certain active working elements, which are dipped in the environment, do not requires start the inversion, since the main role in time of the optical pumping process is given four-photon interaction and the associated amplification of those frequencies components, that result from this interaction, and participating in the PL or generation.